\documentclass{article}

\PassOptionsToPackage{round}{natbib}

 \usepackage[dblblindworkshop, final]{neurips_2025}
 \usepackage{aas_macros}
 \usepackage{CJKutf8}

\usepackage[utf8]{inputenc} 
\usepackage[T1]{fontenc}    
\usepackage{hyperref}       
\usepackage{url}            
\usepackage{booktabs}       
\usepackage{amsfonts}       
\usepackage{nicefrac}       
\usepackage{microtype}      
\usepackage{xcolor}         

\usepackage{graphicx}
\usepackage{amsmath}

\title{Integral Field Unit Spectroscopy with One Fiber}

\author{%
Zehao Peng (\begin{CJK*}{UTF8}{bsmi}彭則皓\end{CJK*})$^{1, 2}$, \ Biprateep Dey$^{1,2}$\thanks{Corresponding author: Biprateep Dey (\texttt{biprateep@pitt.edu})}, \ Chris J.~Maddison$^{1,2}$, \ Joshua S.~Speagle (\begin{CJK*}{UTF8}{gbsn}沈佳士\end{CJK*})$^{1}$ \\ \\
$^{1}$University of Toronto \quad $^{2}$Vector Institute\\
}

\begin{document}

\maketitle

\begin{abstract}
 Integral field unit (IFU) spectroscopy provides spatially resolved spectra across galaxies, offering crucial insights into their evolution. However, its high observational cost limits current IFU datasets to $\sim 10^4$ objects. We present a multi-modal, probabilistic foundation model that predicts high-resolution spectra with calibrated uncertainties at arbitrary spatial locations within a galaxy directly from broadband images. Built on a masked autoencoder framework, our architecture injects fiber positional encodings and redshift aware wavelength encodings, enabling spatially conditioned predictions. Trained on 4.7 million images and single fiber spectroscopic observations from the Dark Energy Spectroscopic Instrument (DESI) survey, our model exploits the natural variance of fiber placements and the morphological self-similarity of galaxies to achieve IFU-like capabilities without any IFU training data. Predicted emission line flux maps match independent IFU observations from the  Mapping Nearby Galaxies at APO (MaNGA) survey, with performance comparable to a supervised baseline trained directly on IFU data. 
\end{abstract}

\section{Introduction} \label{introduction}

Understanding galaxy evolution requires characterizing galaxy properties such as chemical composition, star formation rates, and stellar ages, both at the population level and also within individual galaxies. Spatially resolved mappings of these features can reveal critical variations in stellar populations, gas kinematics, and chemical abundances across galactic discs and bulges. This allows us to investigate the relationship between star formation and galactic structure, as well as the impact of the surrounding environment on local regions within a galaxy.

Large astronomical surveys, such as the Sloan Digital Sky Survey \citep[SDSS;][]{York2000SDSS} and the Dark Energy Spectroscopic Instrument survey \citep[DESI;][]{Dey2019LegacySurvey, DESI2022DESI}, collect broadband images that reveal morphology and single-aperture spectra for tens of millions of objects. These spectra provide detailed population properties, but lack spatial resolution within individual galaxies. Integral field unit (IFU) spectroscopy captures these physical details as spatially resolved spectral datacubes; however, the high cost of allocating numerous optical fibers per target limits contemporary IFU datasets, such as Mapping Nearby Galaxies at APO \citep[MaNGA;][]{Bundy2015Manga}, to roughly $10^4$ galaxies. Therefore, bridging the gap between the statistical volume of these broad surveys and the complete spatio-spectral resolution provided by IFUs is essential for unlocking the full scientific potential of modern astronomical observations.

Previous works have applied advances in self-supervised foundation models to bridge this observational gap by predicting spectra directly from broadband images \citep{Wu2020Spectra, Parker2025AION}. However, all existing frameworks to our knowledge produce a single spectrum per galaxy, typically an integrated or object-center prediction, without exploiting the natural variance of  fiber placement locations present in large spectroscopic data sets, or capitalizing on the morphological self-similarity of galaxies to construct fully resolved spatio-spectral representations. Furthermore, current architectures neither account for heteroscedastic observational uncertainties nor provide calibrated predictive uncertainties, which significantly limits their scientific utility.

In this work, we present a novel uncertainty-aware, multi-modal foundation model based on the self-supervised Masked Autoencoder \citep[MAE;][]{He2021MAE} framework. To effectively bridge the imaging-spectroscopy gap, our architecture enforces cross-modal fusion through masked reconstructions and uniquely incorporates both fiber location and cosmological redshift as positional encodings. This spatial and spectral conditioning establishes a coherent rest-frame latent space and enables spatially resolved predictions of spectra. By actively modeling input measurement and output prediction uncertainties to produce well-calibrated probabilistic reconstructions, our framework synthesizes robust representations that demonstrate powerful generalization capabilities.


\section{Data and Methods}
\label{methods}

\paragraph{Data Sets Used} We cross-match photometry from the DESI Legacy Imaging Survey Data Release~9 (DR9) with spectra from DESI DR1 \citep{DESI2025DR1}, selecting spectroscopically confirmed galaxies and quasars with extinction-corrected $r$-band magnitudes less than $19.5$ and reliable redshift measurements ($\Delta\chi^2 > 40$). For each object, we extract $128 \times 128$ pixel image cutouts across 5 or 6 channels ($g$, $r$, $z$, $W1$, $W2$, plus DR10 $z$-band when available) alongside corresponding spectra spanning 3600--9824 \AA{} (7781 pixels at 0.8 \AA{}/pixel). To enable spatial conditioning and uncertainty modeling, we retain precise on-sky coordinates for both images and spectroscopic fibers, and convert pixel-wise inverse variances into standard measurement errors. We mask missing channels and non-finite measurements. The final curated dataset comprises 4.7 million objects, with 98\% training, 1\% held-out validation for hyperparameter selection, and 1\% held-out testing.
\begin{figure*}
    \centering
    \includegraphics[width=1\linewidth]{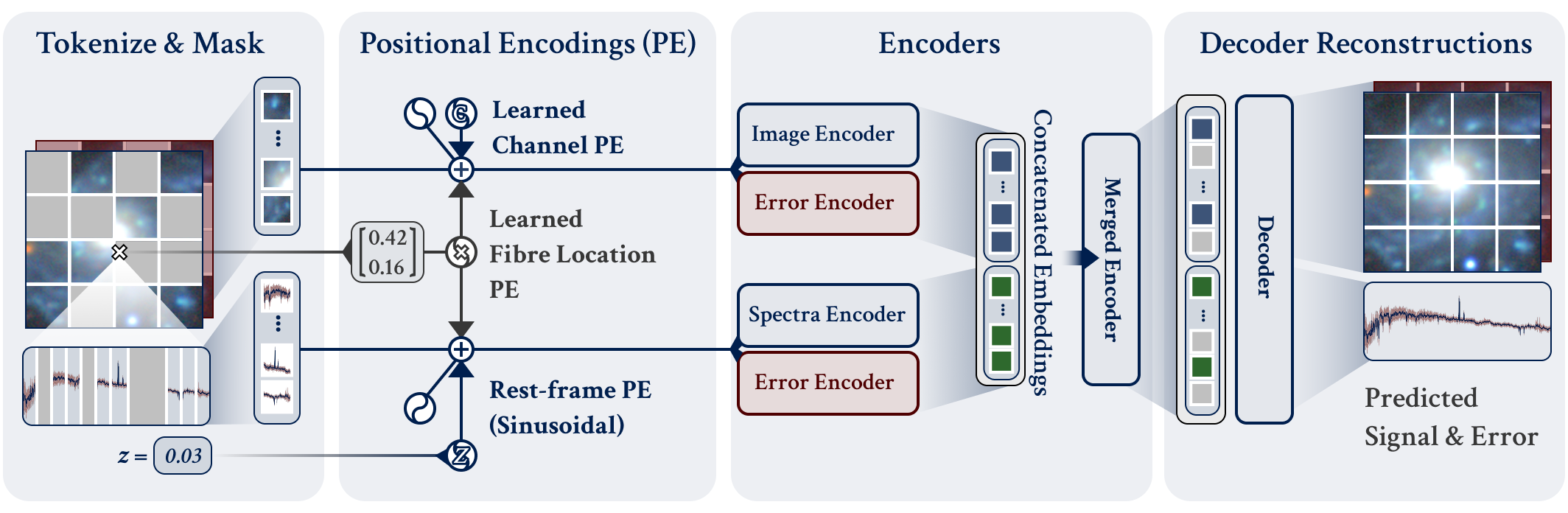}
    \caption{\small \textbf{Schematic representation of our multi-modal Masked Autoencoder architecture.} The model processes six inputs: a galaxy's image, spectrum, their respective measurement errors, the on-sky fiber location and redshift. Following independent tokenization and masking, the image and spectral sequences receive specialized positional encodings (PE). The fiber location is injected into both modalities for spatial conditioning. Spectral tokens additionally incorporate a redshift-informed sinusoidal PE to establish a rest-frame latent space. Finally, these encoded sequences are concatenated and passed through a merged cross-modal encoder, followed by a joint decoder that reconstructs the full images and spectra alongside their predicted uncertainties.}
    \label{fig:model}
    \vspace{-0.4cm}
\end{figure*}

\paragraph{Tokenization and Positional Encoding} Following vision transformers  \citep[ViT;][]{Dosovitskiy2021ViT}, images are divided into $16 \times 16$ pixel patches per channel and spectra into 31-pixel patches, with fluxes and errors tokenized separately and projected to a $256$ dimensional embedding. Channels are tokenized independently to enable direct cross-channel mixing and explicit association between spectral features and broadband morphology. To encourage robust cross-modal representation learning, we independently mask the image and spectral sequences. Image tokens receive a fixed 2D sinusoidal spatial encoding and a learnable channel encoding \citep{Gehring2017positionalembeddings}. We explicitly inject the spectroscopic fiber's on-sky location as a fiber positional encoding by projecting its relative spatial offset from the image center via a learned linear transformation, adding this to each image token. For the spectrum, we directly add the 2D sinusoidal spatial encoding of the fiber's center to all spectral tokens, establishing a spatially conditioned mapping. Because galaxies at different redshifts have their spectra shifted to different observed wavelengths, we use similar sinusoidal encodings to inject the rest-frame wavelengths, enforcing a latent representation that is invariant to redshifts.

\paragraph{Model Architecture} 
Our Transformer-based \citep{Vaswani2017Transformer} masked autoencoder randomly removes tokens from the encoder sequence and is trained to reconstruct the complete data. Masking is applied independently to each modality. To avoid overfitting to a fixed masking regime, we dynamically sample the mask ratio $r \in \{0, 1/14, \dots, 13/14, 1\}$, with $P(r=1)=0.3$ and uniform probability $0.05$ otherwise. This also disproportionately exposes the model to cases where an entire modality is missing, encouraging cross-modal reconstruction. In addition, each ratio is paired with a contiguous chunk size $c \in \{1,2,4,8,16,32,64,128\}$, chosen to avoid pathological masking patterns. Modality-specific encoders (4 layers, 8 heads) process masked sequences separately; their outputs are concatenated and passed through a merged cross-modal encoder (4 layers, 8 heads). For reconstruction, sequences are projected to $512$ dimensions, padded with learned mask tokens, re-injected with positional encodings, and processed by a joint decoder (8 layers, 16 heads) that routes to four feed-forward heads reconstructing image fluxes, image errors, spectral fluxes, and spectral errors. Figure~\ref{fig:model} depicts our architecture and full model details are available in our code repository\footnote{\url{https://github.com/harrypenguin/MultiModal}}.

\paragraph{Loss Function}
Following \citet{Laroche2023scattervae}, we formulate a probabilistic reconstruction objective for heteroscedastic noise. Let $\mathbf{x}_{\text{obs}}$ denote the observed fluxes with pixel-wise known variances $\boldsymbol{\sigma}_{\text{obs}}^2$. We model the true underlying flux as a normal distribution centered on our prediction $\mathbf{x}_{\text{pred}}$ with learned predictive variance $\boldsymbol{\sigma}_{\text{pred}}^2$. This yields a $\chi^2$-type loss function:
$$
\sum \left[ \frac{(\mathbf{x}_{\text{obs}} - \mathbf{x}_{\text{pred}})^2}{\boldsymbol{\sigma}_{\text{obs}}^2 + \boldsymbol{\sigma}_{\text{pred}}^2} + \log(\boldsymbol{\sigma}_{\text{obs}}^2 + \boldsymbol{\sigma}_{\text{pred}}^2) \right]
$$
summed over all image and spectral pixels. The log normalization term penalizes variance inflation where observed data exists. To discourage the model from over-inflating $\boldsymbol{\sigma}_{\text{pred}}^2$ on masked patches where $\mathbf{x}_{\text{obs}}$ is unavailable, we apply a complementary $L_2$ regularization penalty directly to $\boldsymbol{\sigma}_{\text{pred}}^2$ in those regions. During training, we apply random spatial translations to simulate varied off-center fiber placements. The model is optimized using Adam \citep{Kingma2015Adam} with a cosine warmup learning rate schedule.

\section{Results and Discussion}
\label{results}

\begin{figure}[h]
    \centering
    \includegraphics[width=1\linewidth]{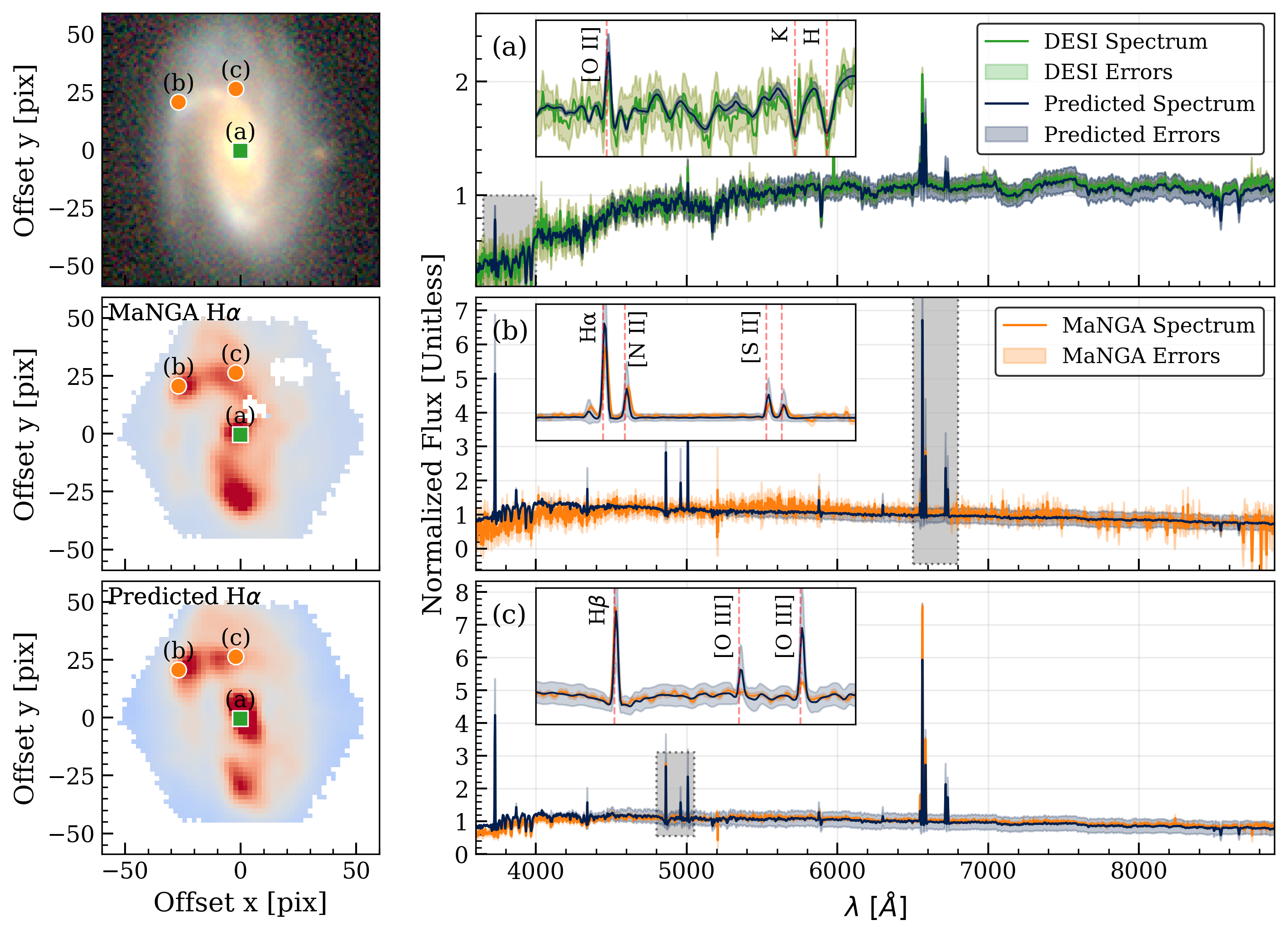}
    \caption{\small \textbf{Predicted spectra at various spatial locations for an example galaxy.} \textbf{Left Column}: (Top) DESI Legacy Survey image of DESI TARGETID 39628444518058848 (MaNGA-ID 9042-9101) at redshift $\sim0.07$. The model is trained to infer spectra based solely on the broadband image, the redshift of the object, and a specified fiber location. (Middle) True H$\alpha$ flux map of the object, derived from the MaNGA IFU datacube. (Bottom) Reconstructed H$\alpha$ flux map derived from spectra predicted by our model. These spectra are generated using only the image and an input fiber location that is systematically scanned across the image at the spatial resolution of the MaNGA IFU. Three spaxels are labelled (a), (b), and (c). Our model qualitatively recovers the H$\alpha$ map and successfully resolves star-forming regions along the galactic arms, denoted by localized hotspots. \textbf{Right Column}: Each panel corresponds to a chosen spaxel, displaying the predicted spectrum and uncertainties in dark blue. Panel (a) compares the model prediction to the available DESI spectrum observed at the central location (green), whereas panels (b) and (c) compare off-center predictions to the corresponding spectra from the MaNGA IFU (orange). The insets zoom into areas of interest to highlight the accurate reconstruction of prominent emission and absorption features. Notably, even though our model was never trained on MaNGA data, it successfully reconstructs off-center spectra and captures the spatially varying physical properties of the galaxy. These predictions trained on DESI data remarkably match a completely independent dataset (up to a normalization constant from flux calibration differences between the two instruments).}
    \label{fig:ifu}
     \vspace{-0.5cm}
\end{figure}

Our model achieves strong reconstruction performance across image and spectral generation tasks over a wide range of masking ratios. Figure~\ref{fig:ifu} (right) demonstrates that the model accurately reconstructs DESI spectra at precise fiber locations, recovering global continuum shapes and local structures, such as oxygen doublets and Ca H and K lines, directly from the broadband image. Because the model explicitly accounts for observational uncertainties, the predicted spectrum represents a denoised inference of the true underlying spectrum.

By systematically scanning the input fiber location across the image at the spatial resolution of the MaNGA IFU, we generate a full synthetic datacube directly from a single broadband image. The predicted spatially resolved emission-line maps and individual spaxel spectra adhere closely to independent MaNGA Data Release 17 ground-truth observations, with slight under-prediction of H$\alpha$ fluxes in select star-forming regions. To quantify performance, we compare our zero-shot predictions against a supervised ridge regression baseline trained on image patches paired with MaNGA H$\alpha$ flux measurements across 470 galaxies. On a held-out test set of 48 galaxies, our foundation model achieves comparable or superior performance to this baseline (Pearson $r \sim 0.71$ for both, and normalized RMSE $1.51$ vs.\ $1.63$ for the baseline). This proves that unsupervised cross-modal pretraining effectively captures the necessary spatio-spectral structures, successfully generalizing to completely unseen data in a zero-shot capacity. 

We further evaluate the calibration of the model's uncertainty estimates on the held-out test set. The distribution of prediction errors weighted by total variance is expected to follow $\mathcal{N}(0,1)$ if the uncertainty estimates are calibrated. Across varying masking ratios, we find that the distribution resembles a Gaussian with a standard deviation between 1.01 and 1.02 for spectra and 0.74 to 0.99 for images, and the median ranges from 0.01 to 0.04 across both modalities. Although the spectral predictions are slightly over-confident and image predictions moderately under-confident, the uncertainties remain well-calibrated overall.

This near-IFU capability reflects a powerful inductive bias learned from the training data distribution. Across 4.7 million DESI observations with naturally varying off-center fiber placements, the model learns that morphologically similar sub-regions (e.g., spiral arms, bulges, and star-forming regions) produce similar spectra regardless of their host galaxy. While this self-similarity assumption is well-grounded for the bulk galaxy population, performance may degrade for morphologically peculiar systems such as ongoing mergers or AGN-dominated galaxies.

Currently, our model ingests redshift as a fixed, uniform input across the spatial dimensions of a given galaxy. This flattens the distribution along the velocity axis and precludes the generation of kinematic maps. This limitation could be addressed by masking the redshift during training and introducing a corresponding reconstruction objective. Additionally, our assumption of Gaussian uncertainty is rigid; future work should explore more flexible probabilistic inference frameworks, such as flow matching or diffusion models.

In conclusion, we demonstrate for the first time the prediction of high-resolution, spatially resolved spectra with calibrated uncertainties at arbitrary locations within a galaxy directly from broadband imaging. By exploiting the natural variance of fiber placements and galactic self-similarity, our model develops near-IFU capabilities using only single-aperture training data. Combined with reliable predictive uncertainties, this approach can significantly expand the scientific yield of current and future photometric surveys to study localized galactic properties. The predictions from our method could enable us to examine specific regions within a galaxy and assess how the environment influences local star formation. Such data sets could also uncover radial gradients in age and metallicity and allow us to study the link between star formation and galactic structure for galaxies without existing IFU data. Furthermore, our method holds substantial potential for broader multi-modal applications in fields such as remote sensing and biology.

\begin{ack}
B. Dey is a postdoctoral fellow at the University of Toronto in the Eric and Wendy Schmidt AI in Science Postdoctoral Fellowship Program, a program of Schmidt Sciences.

The authors advocate for the judicious and ethical use of artificial intelligence in scientific discourse. In accordance with the \href{https://journals.aps.org/authors/ai-based-writing-tools}{American Physical Society’s guidelines} on appropriate use of AI-based writing tools, we confirm that AI assistance was utilized in the preparation of this statement solely to polish, condense, and edit our original ideas. The authors retain full responsibility for the accuracy, integrity, and content of this document.

This research used resources of the National Energy Research Scientific Computing Center, a DOE Office of Science User Facility supported by the Office of Science of the U.S. Department of Energy under Contract No. DE-AC02- 05CH11231 using NERSC award HEP-ERCAP0033572.

Resources used in preparing this research were provided, in part, by the Province of
Ontario, the Government of Canada through CIFAR, and companies sponsoring the Vector Institute (\url{www.vectorinstitute.ai/partnerships/}).

This research used data obtained with the Dark Energy Spectroscopic Instrument (DESI). DESI construction and operations is managed by the Lawrence Berkeley National Laboratory. This material is based upon work supported by the U.S. Department of Energy, Office of Science, Office of High-Energy Physics, under Contract No. DE–AC02–05CH11231, and by the National Energy Research Scientific Computing Center, a DOE Office of Science User Facility under the same contract. Additional support for DESI was provided by the U.S. National Science Foundation (NSF), Division of Astronomical Sciences under Contract No. AST-0950945 to the NSF’s National Optical-Infrared Astronomy Research Laboratory; the Science and Technology Facilities Council of the United Kingdom; the Gordon and Betty Moore Foundation; the Heising-Simons Foundation; the French Alternative Energies and Atomic Energy Commission (CEA); the National Council of Humanities, Science and Technology of Mexico (CONAHCYT); the Ministry of Science and Innovation of Spain (MICINN), and by the DESI Member Institutions: www.desi.lbl.gov/collaborating-institutions. The DESI collaboration is honored to be permitted to conduct scientific research on I’oligam Du’ag (Kitt Peak), a mountain with particular significance to the Tohono O’odham Nation. Any opinions, findings, and conclusions or recommendations expressed in this material are those of the author(s) and do not necessarily reflect the views of the U.S. National Science Foundation, the U.S. Department of Energy, or any of the listed funding agencies.

The Legacy Surveys consist of three individual and complementary projects: the Dark Energy Camera Legacy Survey (DECaLS; Proposal ID \#2014B-0404; PIs: David Schlegel and Arjun Dey), the Beijing-Arizona Sky Survey (BASS; NOAO Prop. ID \#2015A-0801; PIs: Zhou Xu and Xiaohui Fan), and the Mayall z-band Legacy Survey (MzLS; Prop. ID \#2016A-0453; PI: Arjun Dey). DECaLS, BASS and MzLS together include data obtained, respectively, at the Blanco telescope, Cerro Tololo Inter-American Observatory, NSF’s NOIRLab; the Bok telescope, Steward Observatory, University of Arizona; and the Mayall telescope, Kitt Peak National Observatory, NOIRLab. Pipeline processing and analyses of the data were supported by NOIRLab and the Lawrence Berkeley National Laboratory (LBNL). The Legacy Surveys project is honored to be permitted to conduct astronomical research on Iolkam Du’ag (Kitt Peak), a mountain with particular significance to the Tohono O’odham Nation.

NOIRLab is operated by the Association of Universities for Research in Astronomy (AURA) under a cooperative agreement with the National Science Foundation. LBNL is managed by the Regents of the University of California under contract to the U.S. Department of Energy.

This project used data obtained with the Dark Energy Camera (DECam), which was constructed by the Dark Energy Survey (DES) collaboration. Funding for the DES Projects has been provided by the U.S. Department of Energy, the U.S. National Science Foundation, the Ministry of Science and Education of Spain, the Science and Technology Facilities Council of the United Kingdom, the Higher Education Funding Council for England, the National Center for Supercomputing Applications at the University of Illinois at Urbana-Champaign, the Kavli Institute of Cosmological Physics at the University of Chicago, Center for Cosmology and Astro-Particle Physics at the Ohio State University, the Mitchell Institute for Fundamental Physics and Astronomy at Texas A\&M University, Financiadora de Estudos e Projetos, Fundacao Carlos Chagas Filho de Amparo, Financiadora de Estudos e Projetos, Fundacao Carlos Chagas Filho de Amparo a Pesquisa do Estado do Rio de Janeiro, Conselho Nacional de Desenvolvimento Cientifico e Tecnologico and the Ministerio da Ciencia, Tecnologia e Inovacao, the Deutsche Forschungsgemeinschaft and the Collaborating Institutions in the Dark Energy Survey. The Collaborating Institutions are Argonne National Laboratory, the University of California at Santa Cruz, the University of Cambridge, Centro de Investigaciones Energeticas, Medioambientales y Tecnologicas-Madrid, the University of Chicago, University College London, the DES-Brazil Consortium, the University of Edinburgh, the Eidgenossische Technische Hochschule (ETH) Zurich, Fermi National Accelerator Laboratory, the University of Illinois at Urbana-Champaign, the Institut de Ciencies de l’Espai (IEEC/CSIC), the Institut de Fisica d’Altes Energies, Lawrence Berkeley National Laboratory, the Ludwig Maximilians Universitat Munchen and the associated Excellence Cluster Universe, the University of Michigan, NSF’s NOIRLab, the University of Nottingham, the Ohio State University, the University of Pennsylvania, the University of Portsmouth, SLAC National Accelerator Laboratory, Stanford University, the University of Sussex, and Texas A\&M University.

BASS is a key project of the Telescope Access Program (TAP), which has been funded by the National Astronomical Observatories of China, the Chinese Academy of Sciences (the Strategic Priority Research Program “The Emergence of Cosmological Structures” Grant \# XDB09000000), and the Special Fund for Astronomy from the Ministry of Finance. The BASS is also supported by the External Cooperation Program of Chinese Academy of Sciences (Grant \# 114A11KYSB20160057), and Chinese National Natural Science Foundation (Grant \# 12120101003, \# 11433005).

The Legacy Survey team makes use of data products from the Near-Earth Object Wide-field Infrared Survey Explorer (NEOWISE), which is a project of the Jet Propulsion Laboratory/California Institute of Technology. NEOWISE is funded by the National Aeronautics and Space Administration.

The Legacy Surveys imaging of the DESI footprint is supported by the Director, Office of Science, Office of High Energy Physics of the U.S. Department of Energy under Contract No. DE-AC02-05CH1123, by the National Energy Research Scientific Computing Center, a DOE Office of Science User Facility under the same contract; and by the U.S. National Science Foundation, Division of Astronomical Sciences under Contract No. AST-0950945 to NOAO.

\end{ack}







\bibliographystyle{plainnat}
\bibliography{references}

\end{document}